# The Effect of Fluid Properties on the Operation of Thermal Bubble jet

**F.Mobadersani**
*M.Sc. student of mechanical engineering, Iran University of science and technology (IUST)*
F_Mobadersani@yahoo.com

**H.Saffari**
*Assistant professor, department of mechanical engineering, Iran University of science and technology (IUST)*
saffari@iust.ac.ir

**T.Hajilounezhad**
*M.Sc. student of mechanical engineering, University of Tabriz*
thmeen@gmail.com

**P.Kahroba**
*M.Sc. student of mechanical engineering, University of Uroumie*
Pkme61@gmail.com

**Abstract**
Over the last two decades, since explosive boiling has been widely used in industry, research on it has been increased. Thermal bubble jet printer, micro injectors and using in micro medicine for injection are some possible applications.
The operation of thermal bubble jets consist of three stages: 1- Heat transfer process, 2- Bubble formation and growth in the microchannel, and 3-Drop ejection. The effect of fluid properties on the operation of thermal bubble jet is investigated in this paper. Thus the effect of fluid properties is investigated in each above mentioned processes. Eventually, comparing experimental results of drop ejection with the results from simulations, drop properties such as volume and velocity are given.

**Introduction**
Various experiments have been carried out on a thin wire or a very thin microheater to investigate the conditions of explosive bubble formation and growth, where maximum superheat temperature, acoustic pressure due to bubble growth, bubble volume and etc have been investigated [2]-[3]. Our knowledge of explosive vaporization process in microscopic cases on a heater is limited.
Micropumps have various applications like injection of microdrugs, injection of chemical materials, DNA injection, and fuel injection to microengines, and etc. bubble growth in microchannel results in pressure difference inside it which is used to do work. The most promising area of its application is in micropumps. Various experiments have been carried out to investigate the generated pressure difference in microchannel, volume flow rates in both ends of microchannel and bubble volume in microchannel. [2]- [6]
Asai has introduced a new theoretical approach for boiling and bubble growth under the influence of a very high heat flux. He obtained a new analytic approach for boiling properties about boiling possibility in a region of variable temperature with time by means of integrating in terms of time and space. He has proposed an analytical term variable with time for bubble growth which declares the behavior of the thin film of bubble. He has carried out an experiment using methanol with the heat flux of 5-50 MWm$^{-2}$ on a microheater to validate the proposed model, which is in a good harmony with the model results [9].
When the major part of thermal energy appears as to increase bubble volume, the emitted energy while liquid boiling, and the energy released from expanding microbubbles may be utilized to run the micromechanical systems. Effective utilization of the high pressure pulses generated by the explosion of microscopic bubbles may lead to a revolution in thermal micromechanics, where a specific amount of force is required. Explosive boiling, in general, may be generated by an unsteady heat transfer to a liquid by means of a very thin microheater. A very powerful thermal pulse is applied to a very thin layer of liquid to achieve the limit of superheat.
When the heater dimensions are much greater than the thermal boundary layer, heat transfer may be simulated using a one dimensional model. To investigate the effect of fluid properties on the operation of thermal bubble jet, we have divided the whole process into the following three parts in this paper: heat transfer, bubble growth and drop ejection. Then we have investigated the properties of each fluid independently in every three parts. The first part involves solving the one dimensional conduction equation; the second involves bubble growth analysis using Flow3D software package, and finally the last part involves investigation and comparison of the effect of fluid properties on the operation of drop ejection.

**Theoretical Review**
Explosive boiling phenomenon may be considered a homogeneous boiling. Consequently, liquid temperature has to be increased up to homogeneous boiling temperature (that is $300\,^0C$ for water at p=1 atm). Liquid temperature increases significantly due to high heat flux transferred over a short pulse of time (7 $\mu s$ for this work) through conduction (convection is ignored). Since homogeneous nucleation temperature of water is greater than ethanol and methanol, so water nucleation takes place in a longer time. Bubble growth, until it reaches its maximum volume, may be considered in two stages: at early stage pressure (p$_g$) increases abruptly at boiling time which is (1-10) MPa for water, ethanol and methanol. At early stage where cooling effect is not dominant, following relations (A.Asai) may be obtained using claypeyron relation[9]

$$p_v = p_g \exp[-(1+\frac{t}{t_1})(\frac{t}{t_2})^{1/2}]$$

$$t_1 = \frac{3q_h S_h A_l}{2 p_g \rho_g L_g}$$

$$t_2 = \frac{\pi(\alpha_g \beta_g T_g k_l)^2}{4 q_h^2 a_l} \qquad [1]$$

$$\alpha_g \equiv 1 - \frac{\rho_g}{\rho_l}$$

$$\beta_g = \frac{p_g}{\rho_g L_g}$$

Where $p_v$ is vapor pressure, $p_g, \rho_g$ and $L_g$ are gas pressure, density and latent heat at $T_g$, respectively, $\rho_l, a_l, k_l, q_h$ are density, diffusivity, conductivity of liquid, and heat flux, respectively, $S_h$ is heater surface and $A_l$ is the liquid inertance. It has to be noted that equation (1) is just applicable for the early stage where the heat flux transferred to the vapor ($q_v$) is defined as a linear relation. The relation could be used up to $t_3$, where $q_v$ has the maximum value.

At later stage, pressure decreases drastically until it approaches $P_{sat}(T_{amb})$. A.Asai proposed relation may be used at this stage, too:

$$p_v(t) = [p_g - p_{sat}(T_{amb})]\exp[-(\frac{t}{t_e})^\lambda] + p_{sat}(T_{amb}) \qquad [2]$$

Where $\lambda$ is a parameter and $t_e$ is time constant of pressure reduction. The behavior of bubble is just inertial after it reaches its ultimate volume, until it collapses.

**Numerical results**
The operation of thermal bubble jets consists of three phenomena:

1-Heat transfering until boiling
2-Bubble formation and growth
3-Drop ejection

These three phenomena are investigated independently to compare the effect of fluid properties on the operation of thermal bubble jets and the effect of properties is analyzed at each stage

**1-Heat transfers until boiling**
The scheme and geometry of the micropump employed in the present work is shown in Fig.1. Explosive boiling, in general, may be generated with an unsteady heat transfer to liquid by means of a very thin microheater. A very strong heat pulse is applied to the thin layer of liquid to achieve the limit of superheat. The distribution of temperature of liquid in the single phase case may be investigated by the conduction equation considering Cartesian coordinates and ignoring viscosity loss.

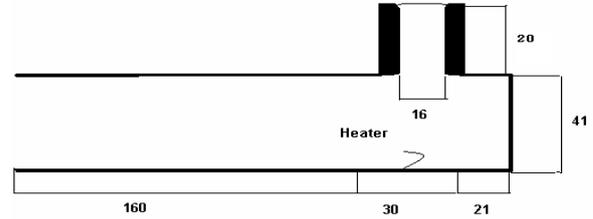

Fig.1 Scheme and geometry of the micropump

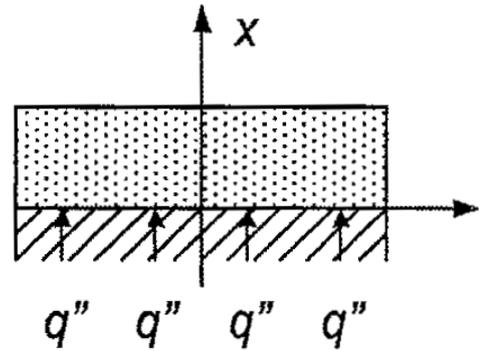

Fig.2 Heat transfer model

$$\frac{\partial T(x,t)}{\partial t} = \frac{1}{\alpha}\frac{\partial^2 T(x,t)}{\partial x^2} \qquad [3]$$

Applying Boundary Conditions

$$\frac{\partial T}{\partial x} = -\frac{q''}{k} \quad q'' = W/A \quad x=0 \quad t>0 \qquad [4]$$

$$T(x,t) = T_0 \quad x \to \infty \quad t>0$$

And Initial Conditions

$$T(x,t) = T_0 \quad t=0 \qquad [5]$$

Conduction equation may have the following solution using separation of variables method when considering constant thermal properties and pressure of liquid.

$$T(x,t) - T_0 = \frac{q''}{k}\int_x^\infty erf(\frac{x'}{\sqrt{4\alpha t}})dx' = \frac{2q''}{k}[(\frac{\alpha t}{\pi})^{1/2}\exp(-\frac{x^2}{4\alpha t}) - \frac{x}{2}erfc(\frac{x}{\sqrt{4\alpha t}})] \quad [6]$$

This equation is solved by Mathematica for ethanol, methanol and water in order to obtain the effect of fluid properties on the heat transfer process until boiling occurs.

The results of solving conduction equation for ethanol, methanol and water are shown in Fig.3, Fig.4 and Fig.5, respectively. Heat flux of 60 MW/m^2 is transferred to the liquid during $6\mu s$ by means of heater, and the liquid is initially at 26 °C.

Analytical results show that the location where explosive boiling starts on the heater is maximum 3nm over the heater [8], hence conduction equation is solved up to this.(x-location). Experimental results also show that maximum temperature that fluid reaches before explosive boiling starts is about 90% of critical temperature of fluid [13].

Thus the conduction equation is solved until 90% of critical temperature of each fluid. Table 1 shows the properties of each fluid used in the conduction equation.

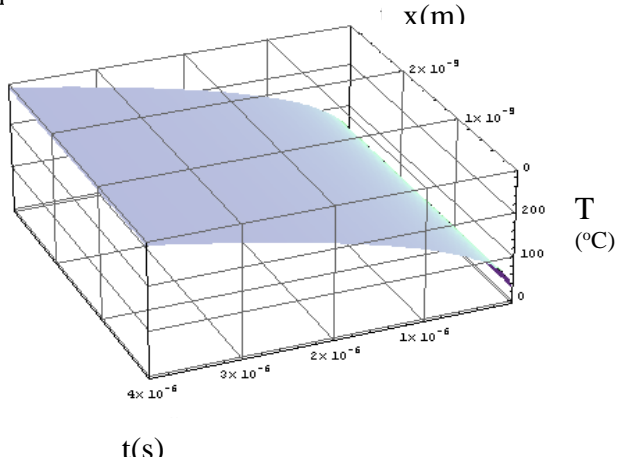

Fig.3 Conduction equation solution on heater for ethanol

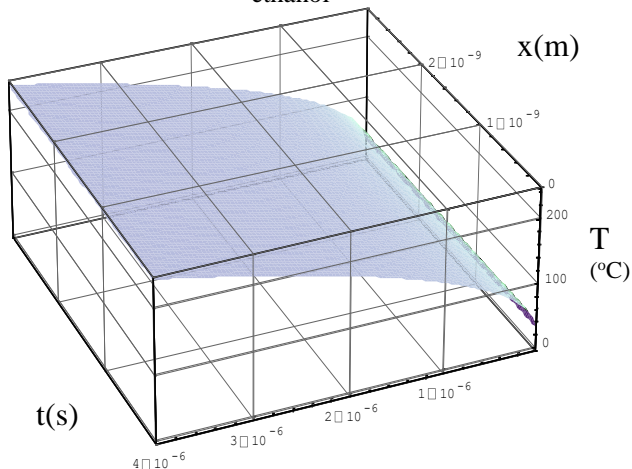

Fig.4 Conduction equation solution on heater for methanol

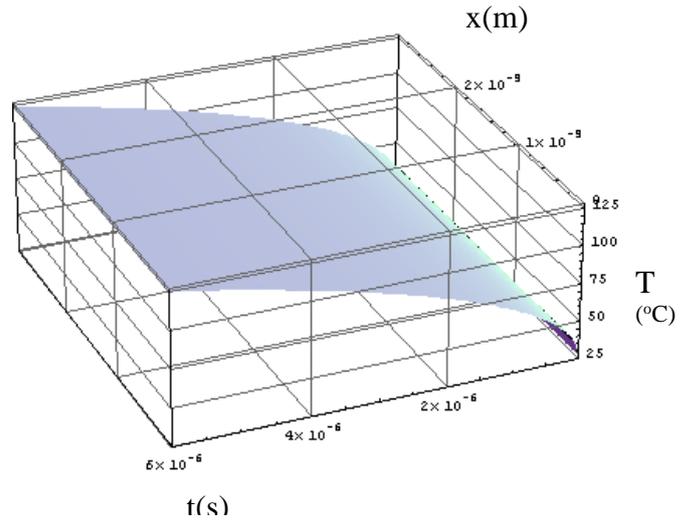

Fig.5 Conduction equation solution on heater for water

As shown in Fig.5 water reaches 125 °C at the end of heat transfer period that is much different from the explosive boiling temperature. Thus under this condition of heat transfer, water may not boil as explosive boiling and boiling may just be as heterogeneous. To investigate the possibility of heterogeneous boiling for water under low heat flux because of cavities, it is supposed in bubble formation stage that water has boiled with this heat flux and the surface of the bubble corresponding this boiling is shown.

| Fluid type | Thermal Conductivity (w/m.K) | Specific Heat (J/kg.K) | Density (Kg/m^3) | Critical Temperature (°C) |
|---|---|---|---|---|
| Water | 0.67 | 4200 | 1000 | 374 |
| Methanol | 0.2 | 2544 | 785 | 240 |
| Ethanol | 0.14 | 2453 | 789 | 241 |

Table1. The properties of each fluid used in the conduction equation

**2-Bubble formation and growth**

Flow3D covers heater surface with a thin layer of fluid vapor at boiling time in order to simulate boiling phenomenon, then starts bubble growth analysis. As shown in Fig. 4 and Fig.6 ethanol and methanol will boil explosively under the heat flux of 60 MW/m^2 at $4\mu s$ and $3.4\mu s$, respectively (these values are somewhat rough). It should be noted that the best means of obtaining boiling time is experimental results, though in absence of them for our case, it is calculated as above mentioned. If the fluid boils explosively, its initial pressure will be drastically high, thus results in the abrupt growth of the bubble. Bubble initial pressure for each fluid is shown in Fig.6. As shown in Fig.6 the initial pressure of water is so small and its bubble dissipates quickly. A relatively high pressure due to cavitation rebound is seen in water curve as shown in Fig.6

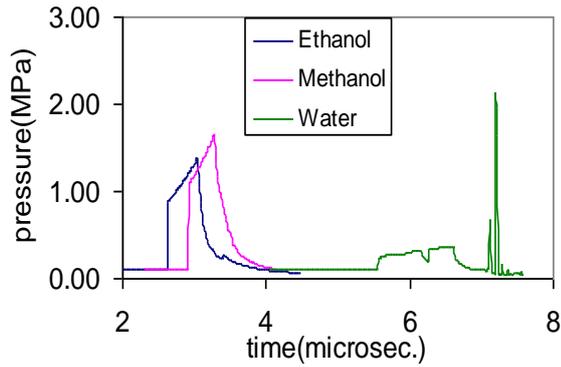

Fig.6 Bubble initial pressure while boiling

Area of the bubble on the surface for water, ethanol and methanol is shown in Fig.7. As it is evident from Fig.7 ethanol bubble has a greater area on the surface than methanol.

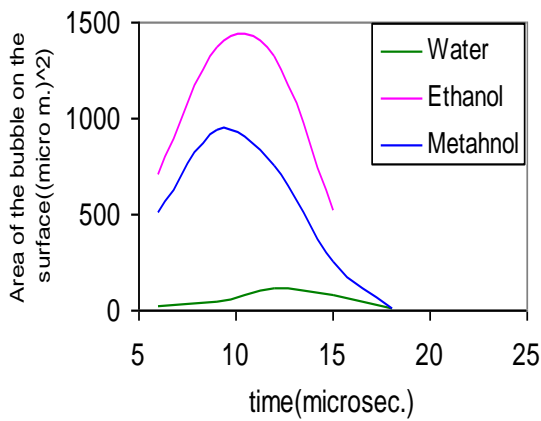

Fig.7 Bubble surface in terms of time

As mentioned previously, water boils heterogeneously under the given heat flux, provided that there are cavities on the bubble heater to start nucleation. Bubble volume is shown in Fig.7 with the assumption of heterogeneous boiling.

**3- Drop ejection**
An experimental model is considered to investigate the effect of fluid properties on drop ejection, then after simulation of drop ejection in this nozzle and comparing with experimental results the effect of fluid properties such as surface tension, viscosity and density is investigated on the operation of drop ejection.
The nozzle of BJ-80 is chosen to compare with experimental results whose geometry is shown in Fig.8. The fluid inside it is some kind of ink of surface tension of 53.8 mN.m$^{-1}$ and viscosity of 4.5mPa.s, and the rest of its properties are the same as water

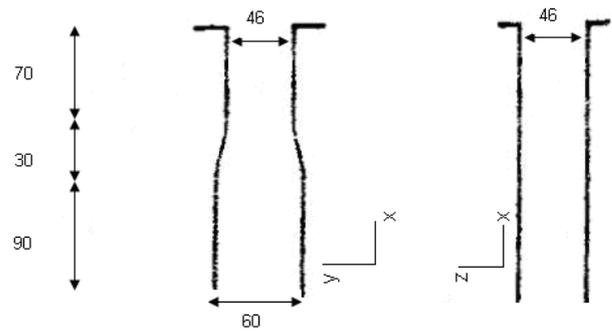

Fig.8 Drop injection Nozzle of BJ-80 printer (all units are $\mu m$)

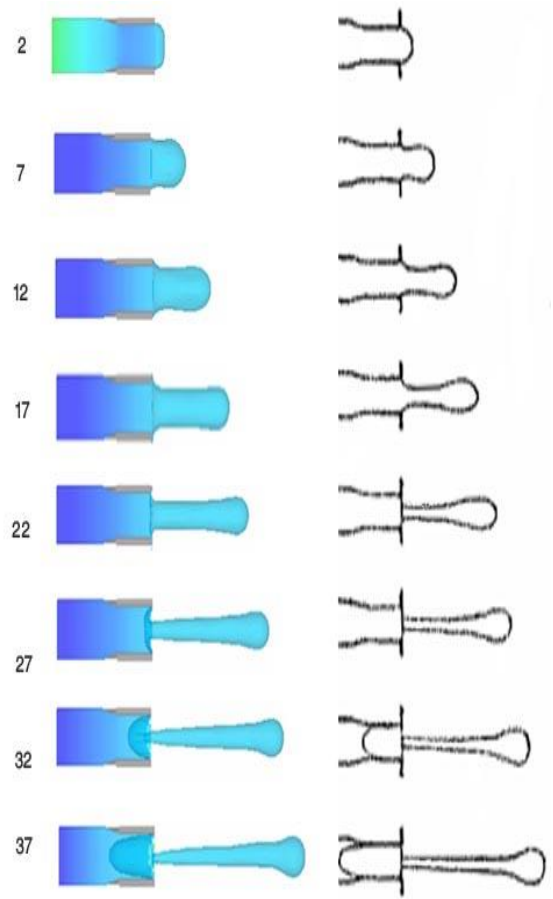

Fig.9 The comparison of experimental results (on the right hand) with simulation results of drop ejection (on the left hand) (unit: $\mu$s)

The effect of surface tension, viscosity and density is shown in Fig.10, Fig.11, Fig.12, respectively.

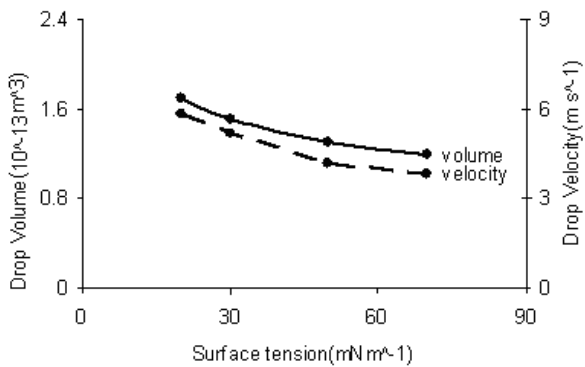

Fig.10 Volume and velocity of the ejected drop as a function of surface tension

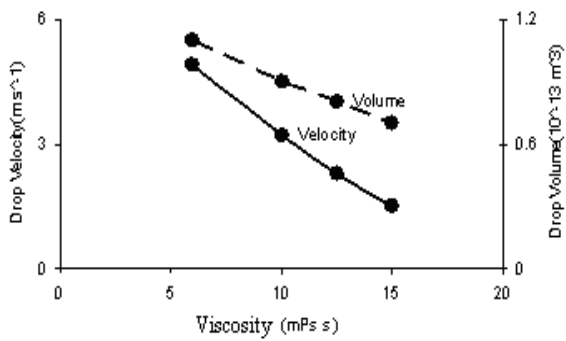

Fig.11 Volume and velocity of the ejected drop as a function of viscosity

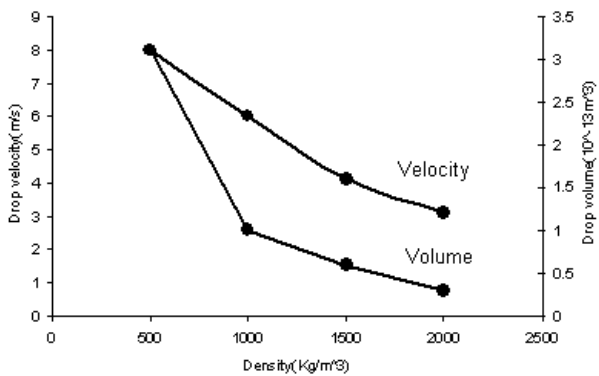

Fig.12 Volume and velocity of the ejected drop as a function of density

The fluid properties are chosen according to experimental results, and the effect of fluid properties, in general, may be determined. The results of simulation are shown in Fig.13 and Fig.14. It is evident from the figures that ethanol drop has a greater volume and velocity than methanol drop and it dissipates sooner.

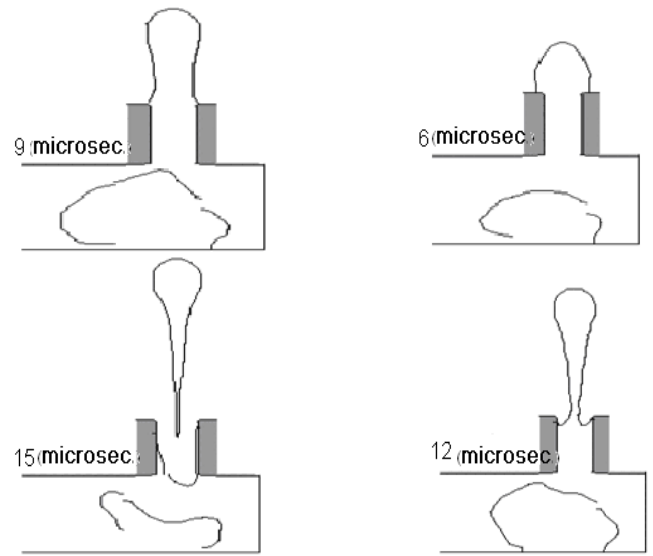

Fig.13 Simulation of ethanol drop ejection

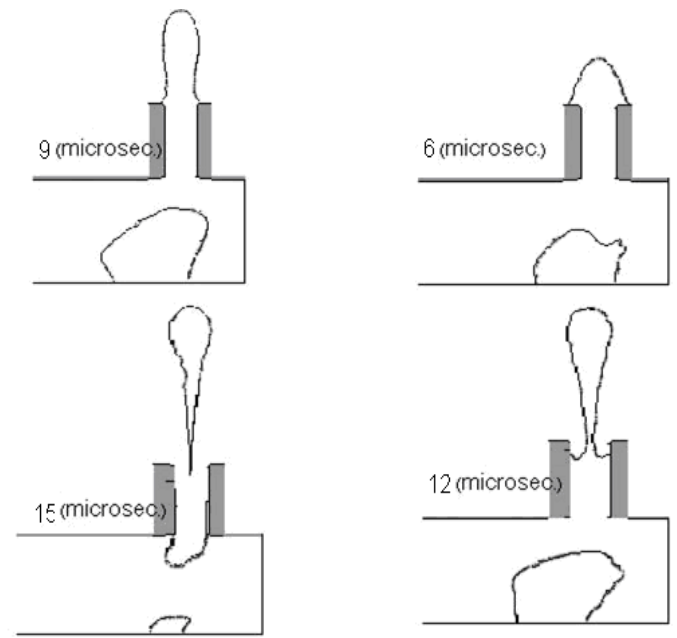

Fig.14 Simulation of methanol drop ejection

The drop properties of ethanol and methanol are compared in table 2.

Table 2 drop properties

| fluid | drop velocity(m/s) | drop volume (m^3*10$^{-13}$) | detaching time ($\mu$ s) |
|---|---|---|---|
| water | Drop is not ejected | | |
| ethanol | 5.7 | 0.25 | 14 |
| methanol | 4.8 | 0.12 | 13 |

## Conclusions

Bubble growth may be utilized provided that its initial pressure is high. Thus heterogeneous boiling can not generate bubbles having the ability to do work. Homogeneous boiling is a process that generates bubbles of high pressure. The effect of fluid properties on heat transfer process is investigated primarily in this paper. It is obvious that the temperature of ethanol reaches the homogeneous boiling temperature more quickly than of methanol, thus, ethanol bubble forms sooner, however, the temperature of water under the given heat flux will not reach the homogeneous boiling temperature. Investigating drop ejection, it is evident that the volume and velocity of the bubble decreases as the viscosity, surface tension and density increases.

## Numenclature

| Symbol | Description |
|---|---|
| $T$ | Temperature |
| $p_v$ | vapor pressure |
| $L_g$ | latent heat |
| $S_h$ | heater surface |
| $P_{sat}(T_{amb})$ | Vapor pressure at ambient pressure |
| $q''$ | Heat flux |
| $\lambda$ | A parameter |
| $\alpha$ | Diffusivity |
| $\sigma$ | Surface tension |
| $\mu_l$ | Viscosity |